\documentstyle[12pt]{article}

\newcommand{\beq}{\begin{eqnarray}}
\newcommand{\ene}{\end{eqnarray}}

\newcommand{\F}{\noindent}

\newcommand{\q}{\quad}

\newcommand{\MP}{\medskip}
\newcommand{\BP}{\bigskip}
\newcommand{\HH}{{\cal H}}

\newcommand{\eq}[1]{(\ref{#1})}

\begin{document}

\rightline{KIMS-1999-10-17}
\rightline{gr-qc/9910081}
\BP

\vskip12pt

\vskip8pt

\Large

\begin{center}
{\bf A possible solution for the non-existence of time}
\vskip10pt

\normalsize
Hitoshi Kitada

Department of Mathematical Sciences

University of Tokyo

Komaba, Meguro, Tokyo 153-8914, Japan

e-mail: kitada@kims.ms.u-tokyo.ac.jp

http://kims.ms.u-tokyo.ac.jp/

\vskip8pt

November 12, 1999

\end{center}
\MP

\vskip10pt

\leftskip=24pt
\rightskip=24pt

\small

\F
{\bf Abstract.} 
A possible solution for the problem of non-existence
 of universal time is given by utilizing G\"odel's
incompleteness theorem \cite{G}.

\MP

\leftskip=0pt
\rightskip=0pt

\normalsize

\vskip 8pt

In a recent book \cite{Barbour}, Barbour presented
 a thought that time is an illusion, by noting that
 Wheeler-DeWitt equation yields the non-existence of
 time, whereas time around us seems to be flowing.
 However, he does not appear to have a definite idea 
 or formal way to actualize his thought. In the present
 note I present a concrete way to resolve the problem
 of the non-existence of time, which is partly a
 reminiscence of my works \cite{K1}, \cite{K2},
 \cite{Ki-Fl1}, \cite{Ki-Fl2}.

\BP

\F
{\bf 1. Time seems not to exist}
\BP

According to equation (5.13) in Hartle \cite{CP},
 the non-existence of time would be expressed by
 an equation:
\beq
H \Psi=0.\label{1}
\ene
Here $\Psi$ is the ``state" of the universe
belonging to a suitable Hilbert space $\HH$, and
$H$ 
denotes the total Hamiltonian of the universe defined
in $\HH$. This equation implies that there is no
 global time of the universe, as the state $\Psi$
 of the universe is an eigenstate for the total
 Hamiltonian $H$, and therefore does not change.
One might think that this implies the non-existence
 of local time because any part of the universe is
 described by a part of $\Psi$. Then we have no time,
 in contradiction with our observations. This is a restatement
 of the problem of time, which is a general problem to
 identify a time coordinate while preserving the
 diffeomorphism invariance of General Relativity.
 In fact, equation \eq{1}
 follows if one assumes the existence of a preferred
 foliating family of spacelike surfaces in spacetime
 (see section 5 of \cite{CP}).

We give a solution in the paper to this problem that
 on the level of the total universe, time does not exist,
 but on the local level of our neighborhood, time does exist.

\newpage

\BP

\F
{\bf 2. G\"odel's theorem}
\BP

Our starting point is the incompleteness theorem proved by
 G\"odel \cite{G}. It states that any consistent
 formal theory that can
 describe number theory includes an infinite number of undecidable
 propositions. The physical world includes at least
 natural numbers, and it is described by a system of words, which
 can be translated into a formal physics theory. The theory of
 physics, if consistent, 
 therefore includes an undecidable proposition, i.e. a proposition
 whose correctness cannot be known by human beings until one finds
 a phenomenon or observation that supports the proposition or
 denies
 the proposition. Such propositions exist infinitely according to
 G\"odel's theorem. Thus human beings, or any other finite entity,
 will never be able to reach a ``final" theory that can express
 the totality of the phenomena in the universe.

Thus we have to assume that any human observer sees a part
 or subsystem $L$ of the universe and never gets the total
 Hamiltonian $H$ in \eq{1} by his observation. Here the
 total Hamiltonian $H$ is an {\it ideal} Hamiltonian
that might be gotten by ``God." In other words, a consequence
 from G\"odel's theorem is that the Hamiltonian that an
 observer assumes with his observable universe
 is a part $H_L$ of $H$. Stating explicitly, 
the consequence from G\"odel's theorem is the
 following proposition
\beq
H=H_L+I+H_E,\q H_E\ne 0,\label{2}
\ene
where $H_E$ is an unknown Hamiltonian describing
 the system $E$ exterior to the realm of the observer,
 whose existence, i.e. $H_E\ne 0$, is assured by G\"odel's
 theorem. This unknown system $E$ includes
all that is unknown to the observer. 
E.g., it might contain particles which
 exist near us but have not been discovered yet,
 or are unobservable for some reason at the time of
 observation.
The term $I$ is an unknown interaction between
 the observed system $L$ and the unknown system $E$. 
Since the exterior system $E$ is assured to exist
by G\"odel's theorem, the interaction $I$ does not vanish:
 In fact assume $I$ vanishes. Then the observed system $L$
 and the exterior system $E$ do not interact, which is
 the same as that the exterior system $E$ does not exist
 for the observer. This contradicts that the observer 
 is able to construct a proposition by G\"odel's
 procedure (see section 5 and \cite{G}) that proves
 $E$ exists. By the same reason, $I$ is not
 a constant operator:
\beq
I \ne \mbox{constant operator}.\label{3}
\ene
For suppose it is a constant operator. Then
 the systems $L$ and $E$ do not change no matter how far or
 how near they are located because the interaction
 between $L$ and $E$ is a constant operator.
 This is the same situation as that the interaction does not
 exist, thus reduces to the case $I=0$ above.

We now arrive at the following observation:
For an observer, the observable universe is a part $L$ 
of the total universe and it looks as though it follows the
 Hamiltonian $H_L$, not following the total Hamiltonian $H$.
 And the state of the system $L$ is described by a part
 $\Psi(\cdot,y)$ of the state $\Psi$ of the total universe,
 where $y$ is an unknown coordinate of system $L$ inside
 the total universe, and $\cdot$ is the variable controllable
 by the observer, which we will denote by $x$.
\BP

\newpage

\F
{\bf 3. Local Time Exists}
\BP

Assume now, as is usually expected, that there is no
 local time of $L$, i.e. that the state $\Psi(x,y)$
 is an eigenstate of the local Hamiltonian $H_L$ for
 some $y=y_0$ and a real number $\mu$:
\beq
H_L\Psi(x,y_0)=\mu\Psi(x,y_0).\label{4}
\ene
Then from \eq{1}, \eq{2} and \eq{4} follows that
\beq
&&0=H\Psi(x,y_0)
=H_L\Psi(x,y_0)+I(x,y_0)\Psi(x,y_0)+H_E\Psi(x,y_0)\nonumber\\
&&\ \hskip5pt=(\mu+I(x,y_0))\Psi(x,y_0)+H_E\Psi(x,y_0).\label{5}
\ene
Here $x$ varies over the possible positions of the particles
 inside
 $L$. On the other hand, since $H_E$ is the Hamiltonian
 describing the system $E$ exterior to $L$, it does not
 affect the variable $x$ and acts only on the variable $y$.
 Thus $H_E\Psi(x,y_0)$ varies as a bare function $\Psi(x,y_0)$
 insofar as the variable $x$ is concerned.
Equation \eq{5} is now written: For all $x$
\beq
H_E\Psi(x,y_0)=-(\mu+I(x,y_0))\Psi(x,y_0).\label{6}
\ene
As we have seen in \eq{3}, the interaction $I$ 
is not a constant operator and varies when $x$
 varies\footnote[2]{Note that G\"odel's theorem
 applies to any fixed $y=y_0$ in \eq{3}. Namely,
 for any position $y_0$ of the system $L$ in the
 universe, the observer must be able to know
 that the exterior system $E$ exists because
 G\"odel's theorem is a universal statement
 valid throughout the universe.
 Hence $I(x,y_0)$ is not a constant operator
 with respect to
 $x$ for any fixed $y_0$.},
 whereas the action
 of $H_E$ on $\Psi$ does not.
 Thus there is a nonempty set of points $x_0$
 where $H_E\Psi(x_0,y_0)$ and $-(\mu+I(x_0,y_0))\Psi(x_0,y_0)$
 are different, and \eq{6} does not hold at such points
 $x_0$. If $I$ is assumed to be continuous in the variables
 $x$ and $y$, these points $x_0$ constitutes a set of
 positive measure. This then implies that our assumption
 \eq{4} is wrong. Thus a subsystem $L$ of the universe cannot
 be a bound state with respect to the observer's Hamiltonian
 $H_L$. This means that the system $L$ is observed as
 a non-stationary system, therefore there must be observed
 a motion inside the system $L$. This proves that the
 ``time" of the local system $L$ {\it exists for the
 observer} as a measure of motion, whereas the total
 universe is stationary and does not have ``time."

\BP

\F
{\bf 4. A refined argument}
\BP

To show the argument in section 3 more explicitly,
we consider a simple case of
$$
H=\frac{1}{2}\sum_{k=1}^N
h^{ab}(X_k)p_{ka} p_{kb}+V(X).
$$
Here $N$ $(1\le N\le \infty)$ is the number of particles
 in the universe, $h^{ab}$ is a three-metric, 
$X_k\in R^3$ is the position of the $k$-th particle, 
$p_{ka}$ is a functional derivative corresponding to
 momenta of the $k$-th particle, and
$V(X)$ is a potential. The configuration
 $X=(X_1,X_2,\cdots,X_N)$ of total particles is decomposed
 as $X=(x,y)$ accordingly to if the $k$-th particle
 is inside $L$ or not, i.e. if the $k$-th particle is
 in $L$, $X_k$ is a component of $x$ and if not it is
 that of $y$. $H$ is decomposed as follows:
$$
H=H_L+I+H_E.
$$
Here $H_L$ is the Hamiltonian of a subsystem $L$ that
 acts only on $x$, $H_E$ is the Hamiltonian describing the
 exterior $E$ of $L$ that acts only on $y$, and
$I=I(x,y)$ is the interaction between the systems $L$ and $E$.
 Note that $H_L$ and $H_E$ commute.

\BP

\noindent
{\bf Theorem.}  \ Let $P$ denote the eigenprojection
onto the space of all bound states of $H$.
Let $P_L$ be the eigenprojection for $H_L$. Then we have
\begin{equation}
(1-P_L)P \ne 0,\label{7}
\end{equation}
unless the interaction $I=I(x,y)$ is a constant with
 respect to $x$ for any $y$.
\MP

\noindent
{\bf Remark.} In the context of the former part,
 the theorem implies the following: 
$$
(1-P_L)P \HH \ne \{ 0 \},
$$
where $\HH$ is a Hilbert space consisting of all
 possible states $\Psi$ of the total universe.
 This relation implies that there is a vector $\Psi\ne 0$
 in $\HH$ which satisfies $H\Psi=\lambda \Psi$ for
 a real number $\lambda$ while $H_L \Phi \ne \mu \Phi$
 for any real number $\mu$, where $\Phi=\Psi(\cdot,y)$
 is a state vector of the subsystem $L$ with an appropriate
 choice of the position $y$ of the subsystem.
\BP

\noindent
{\it Proof} of the theorem. 
Assume that \eq{7} is incorrect. Then we have
$$
P_LP=P.
$$
Taking the adjoint operators on the both sides, we then have
$$
PP_L=P.
$$
Thus $[P_L,P] = P_LP - PP_L = 0$.
 But in generic this does not hold because
$$
[H_L,H] = [H_L, H_L+I+H_E] = [H_L,I]\ne 0,
$$
unless $I(x,y)$ is equal to a constant with respect to $x$.
 Q.E.D.

\BP

\F
{\bf 5. Conclusion}
\BP

G\"odel's proof of the incompleteness theorem relies on the
 following type of proposition $P$ insofar as concerned
 with the meaning:
\beq
P\equiv \mbox{``}P\mbox{ cannot be proved."}\label{8}
\ene
Then if $P$ is provable it contradicts $P$ itself, and if $P$ is not
 provable, $P$ is correct and seems to be provable. Both cases lead
 to contradiction, which makes this kind of proposition undecidable
 in a given consistent formal theory.

This proposition reminds us of the following type of self-referential
 statement:
\beq
\mbox{A person $P$ says ``I am telling a lie."}\label{9}
\ene
The above statement and proposition $P$ in \eq{8} are non-diagonal
 statements
 in the sense that both deny themselves. Namely the core of
 G\"odel's theorem is in proving the existence of non-diagonal
 ``elements" (i.e. propositions) in any formal theory that includes
 number theory. Assigning the so-called G\"odel number to each
 proposition in number theory, G\"odel constructs such
 propositions
 in number theory by a diagonal argument, which shows that any
 consistent formal
 theory has a region exterior to the knowable world.

On the other hand, what we have deduced from G\"odel's theorem in
 section 2 is that the interaction term $I$ is not a constant operator.
 Moreover the argument there implies that $I$ is not diagonalizable
 in the following decomposition of the Hilbert space $\HH$:
\beq
\HH=\int^\oplus \HH_L(\lambda)d\lambda\otimes
\int^\oplus \HH_E(\mu)d\mu,\label{10}
\ene
where the first factor on the RHS is the decomposition of
 $\HH$ with respect to the spectral representation of $H_L$,
 and the second is the one with respect to that of $H_E$.
 In this decomposition, $H_0=H_L+H_E$ is decomposed as a diagonal
 operator:
$$
H_0=H_L\otimes I_E+I_L\otimes H_E=\int^\oplus \lambda d\lambda\otimes I_E
+
I_L \otimes\int^\oplus \mu d\mu,
$$
where $I_L$ and $I_E$ denote identity operators in respective
 factors in \eq{10}. To see that $I$ is not diagonalizable in
 the decomposition \eq{10}, assume contrarily that $I$ is
 diagonalizable with respect to \eq{10}.
 Then by spectral theory of selfadjoint operators, $I$ is
 decomposed as $I=f(H_L)\otimes I_E+I_L\otimes g(H_E)$
 for some functions $f(H_L)$ and $g(H_E)$ of $H_L$ and $H_E$.
 Thus the total Hamiltonian $H$ is also diagonalizable and
 written as:
$$
H=H_0+I=(H_L+f(H_L))\otimes I_E+I_L\otimes(H_E+g(H_E)).
$$
Namely the total Hamiltonian $H$ is decomposed into a sum of
 mutually independent operators in the decomposition of the
 total system into the observable and unobservable systems $L$
 and $E$. This means that there are no interactions between $L$
 and $E$, contradicting G\"odel's theorem as in section 2.
 Therefore $I$ is not diagonalizable with respect to the direct
 integral decomposition \eq{10} of the space $\HH$.

Now a consequence of G\"odel's theorem in the context
 of the decomposition of the total universe into observable and
 unobservable systems $L$ and $E$ is the following:
\begin{quotation}
\F
In the spectral decomposition \eq{10} of $\HH$ with respect to a
 decomposition of the total system into the observable and
 unobservable ones, $I$ is non-diagonalizable. In particular
 so is the total Hamiltonian $H=H_L+I+H_E$.
\end{quotation}
Namely G\"odel's theorem yields the existence of non-diagonal
 elements in the spectral representation of $H$ with respect to
 the decomposition of the universe into observable and
 unobservable systems. The existence of non-diagonal
 elements in this decomposition is the cause that the
 observable state $\Psi(\cdot,y)$ is not a stationary state
 and local time arises, and that decomposition is inevitable by
 the existence of the region unknowable to human beings.

 From the standpoint of the person $P$ in \eq{9}, his universe needs
 to proceed to the future for his statement to be decided true or
 false; the decision of which requires his system to have infinite ``time."
 This is due to the fact that his self-contradictory statement does
 not give him satisfaction in his own world and forces him
 to go out to the region exterior to his universe.
 Likewise, the interaction $I$ in the decomposition above
 forces the observer to anticipate the existence of a
 region exterior to his knowledge. In both cases the unbalance
 caused by the existence of an exterior region yields time.
 In other words, time is an indefinite desire to reach the
 balance that only the universe has.

\BP

\F
{\bf Acknowledgements.} I wish to express my appreciation to
 the members of Time Mailing List at http://www.kitada.com/
 for giving me the opportunity to consider the present problem.
 Special thanks are addressed to Lancelot R. Fletcher, Stephen
 Paul King, Benjamin Nathaniel Goertzel, Bill Eshleman,
 Matti Pitkanen, whose stimulating discussions with me on
 the list have led me to consider the present problem. 
 I especially thank Stephen and Bill for their comments 
 on the earlier drafts to improve my English and descriptions.


\end{document}